\documentclass[conference, 10pt]{IEEEtran} 
\IEEEoverridecommandlockouts

\usepackage{amsmath,amsfonts,bm,amssymb,psfrag,ifthen,color,subfigure}
\usepackage{mathrsfs}

\usepackage[dvips]{epsfig}
\usepackage[dvips]{graphicx}
\usepackage{amsmath,amsfonts,bm,amssymb,psfrag,ifthen,color,subfigure}
\allowdisplaybreaks[1]
\usepackage{algpseudocode}
\usepackage{algorithm}
\usepackage{algorithmicx}
\usepackage{ifthen}
\usepackage{setspace}
\usepackage{cite}
\usepackage{url}
\usepackage{longtable}
\usepackage{stfloats}  
\usepackage{graphics,booktabs,color,subfigure}
\usepackage{float}
\usepackage{array}
\usepackage{authblk}

\newcommand{\st}{{\rm s.t.}}

\usepackage{listings}
\usepackage{stfloats}

\floatname{algorithm}{Algorithm}

\begin{document}

\title{Rate Splitting Multiple Access for Simultaneous Lightwave Information and Power Transfer}

  \author{Zhengqing~Qiu and Yijie~Mao \\
  School of Information Science and Technology, ShanghaiTech University, Shanghai 201210, China\\
  Email: \{qiuzhq2023,maoyj\}@shanghaitech.edu.cn
    \thanks{This work has been supported in part by the National Nature Science Foundation of China under Grant 62201347; and in part by Shanghai Sailing Program under Grant 22YF1428400. (\textit{Corresponding author: Yijie Mao})
   }
   }

\maketitle
\begin{abstract}
This paper initiate the application of rate splitting multiple access (RSMA) for 
simultaneous lightwave information and power transfer (SLIPT), where users require to decode information and harvest energy. 
We focus on a time-splitting (TS) mode where information decoding and energy harvesting are separated in two different phases. 
Based on the proposed system model, we design a constrained-concave-convex programming (CCCP) algorithm to solve the optimization problem of maximizing the worst-case rate among users subject to the harvested energy constraint at each user.
Specifically, the proposed algorithm exploits transformation of the bilinear function, semidefinite relaxation (SDR), CCCP, and a penalty method to effectively deal with the non-convex constraints and objective function. 
Numerical results show that our proposed RSMA-aided SLIPT outperforms the existing baselines based on space-division multiple access (SDMA) and non-orthogonal multiple access (NOMA).

\end{abstract}


\section{introduction}

With the popularity of intelligent terminals and the rise of Internet of Things (IoT) technology, the demand for wireless spectrum resources and data services is constantly increasing.
To meet the growing demand for massive data in 6G indoor networks, visible light communication (VLC) offers various advantages such as a wide,  permission-free spectrum range and high energy efficiency, making it a promising technology for 6G.
Meanwhile, in response to the growing demand for energy-efficient solutions in VLC, the simultaneous lightwave information and power transfer (SLIPT) technique is proposed in \cite{Dia_2018_SLIPT}. By enabling simultaneous transfer of both information and energy to users over the same visible light spectrum, SLIPT provides an efficient solution to satisfy both communication and energy demands. It is particularly beneficial for IoT applications, where devices typically require continuous energy supply and data connectivity.

Despite the above advantages, VLC and SLIPT are also constrained by the limited modulation bandwidth of existing available light-emitting diodes (LEDs).
To overcome this challenge, designing effective multi-access (MA) schemes is a promising research direction for VLC.
Recently, rate-splitting multiple access (RSMA) has been proposed \cite{Mao_2022_rsma_survey} and widely studied in radio frequency (RF) communications. 
It has shown to be superior to many existing MA schemes such as space-division multiple access (SDMA) and power-domain non-orthogonal multiple access (NOMA) \cite{Mao_2022_rsma_survey,Mao_EURASIP_2018}. 
These benefits motivate the exploration of RSMA in VLC.
In \cite{Naser_2022,Tao_ICC_2020,2024_ris_vlc_rsma}, RSMA was investigated in multi-LED VLC systems.
RSMA is shown to achieve higher spectral efficiency than conventional MA schemes such as NOMA and SDMA in these works.
However, the potential benefits of RSMA in SLIPT systems are still in the early stages of exploration.
To the best of the authors' knowledge, only two studies \cite{2024_Girdher_rsma_slipt,2024_guo_slipt_rsma} focused on this topic.
These two works are inspired by the prior research on RSMA for simultaneous wireless information and power transfer (SWIPT) in RF communications \cite{2019_mao_rsma_swipt}, where RSMA is shown to achieve better rate-energy region compared with NOMA and SDMA.
When moving to the visible light spectrum, \cite{2024_Girdher_rsma_slipt} shows that RSMA offers higher harvested energy under certain QoS rate constraints, while \cite{2024_guo_slipt_rsma} demonstrates that RSMA achieves higher spectral efficiency under harvested energy constraints at each user, compared to NOMA and SDMA. 
Although these works have certain merits, the limitations cannot be ignored.


Firstly, for rate expressions, these works all utilize the classic Shannon capacity, which is not suitably applicable to VLC systems due to the unique characteristics inherent in VLC. 
Therefore, specialized capacity formulations and analyses are necessary for VLC systems.
In \cite{S_Ma_2017}, a closed-form lower bound for the achievable rate is presented in VLC while considering both electrical and optical power constraints.
Secondly, these works only consider using photodiodes (PDs) as receivers, where the alternating current (AC) and direct current (DC) components in the received signal can be directly separated by employing two disjoint circuits \cite{2024_guo_slipt_rsma}.
However, the major drawback of PDs is that they require external circuitry and power to operate \cite{wang_2015_solar_panel_design}.
To address these limitations, a solar panel can be used to directly replace PD, converting the optical signal into an electrical signal without the need for an external power supply.
The employment of solar panels can also further simplify the receiver circuitry.

Inspired by the advantages of RSMA in RF and VLC, and the limitations of existing studies on RSMA for SLIPT, this work initiates an investigation into a novel RSMA-aided SLIPT with solar panel-based receiver structures.
The primary contributions of this paper are summarized as:
 \begin{itemize}
    \item 
    We propose a novel RSMA-aided SLIPT system in which a single solar panel serves as the sole receiver for both information and energy harvesting.
    To implement the proposed RSMA for SLIPT, we adopt a time-splitting (TS) mode, which divides the total transmission time block into two phases respectively for communication and energy harvesting.
    To the best of our knowledge, this is the first work that investigates RSMA-aided SLIPT  with a practical solar panel-based receiver structure.
    

    \item 
    Based on the proposed RSMA-aided SLIPT, we derive a close-form lower bound of achievable rates for RSMA-based information decoding (ID).
    And we formulate a beamforming design optimization problem to maximize the worst-cast rate among users, namely the max-min fairness (MMF) rate,  while ensuring that the harvested energy meets a lower bound for each user.

    \begin{figure*}[ht]
      \centering
	\includegraphics[width=16cm]{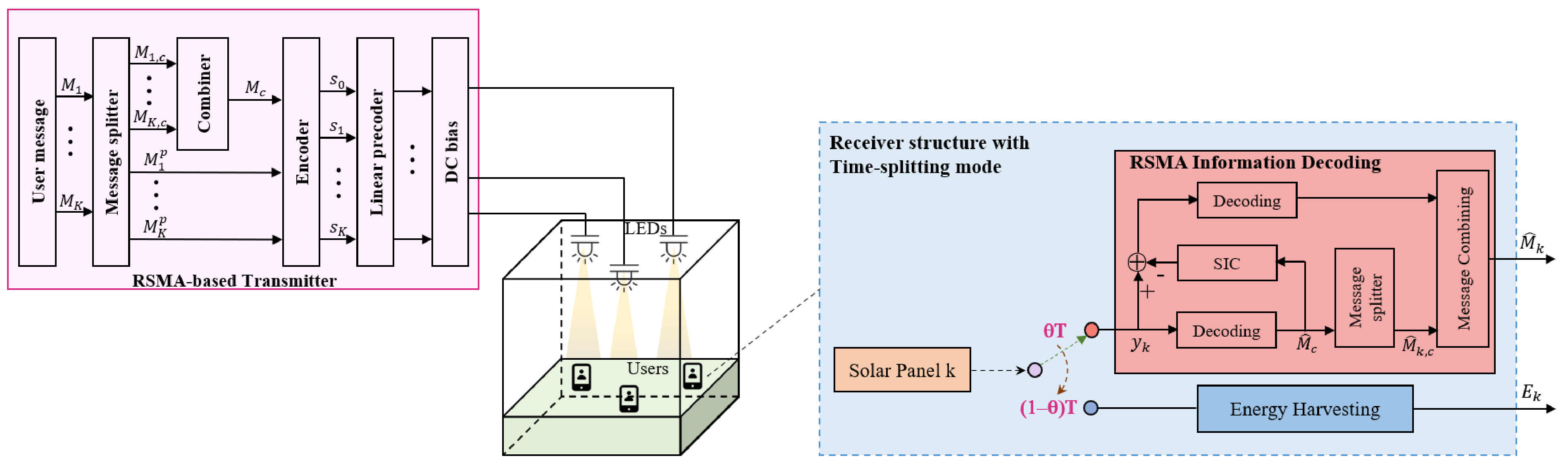}
          \caption{The proposed RSMA-aided multi-user MISO SLIPT based on the TS mode.}
  \label{system_model} 
    \end{figure*}

    \item 
    To address this non-convex MMF rate problem, we propose an efficient optimization algorithm based on the bilinear function transformation, semidefinite relaxation (SDR), constrained-concave-convex programming (CCCP), as well as a penalty method to handle the rank-one constraint associated with SDR.
    Numerical results show that the proposed  RSMA-aided SLIPT enhanced MMF rate compared to conventional SDMA and NOMA-based SLIPT.

  \end{itemize}

\section{system model and problem formulation}
As depicted in Fig. \ref{system_model}, a downlink RSMA-aided multi-user multiple-input single-output (MISO) SLIPT based on the TS mode is considered.
$N$ LEDs indexed by $\mathcal{N}=\{1,\ldots,N\}$ are equipped at the transmitter and they simultaneously serve $K$ users indexed by $\mathcal{K}=\{1,\ldots,K\}$. All users require both information and energy from the lightwave.

\subsection{Channel Model}
The optical wireless channel is predominantly influenced by the line-of-sight (LOS) link, rendering diffuse links negligible. 
Specifically, we represent the channel vector of user $k$ as ${{\mathbf{h}}_k}   = {\left[ {{h_{k,1}}, \ldots ,{h_{k,N}}} \right]^T} \in \mathbb{R}^{N \times 1}$, where the channel $h_{k,n}$ between the $n$th LED and user $k$.
Following the Lambertian model, $h_{k,n}$ is given by
\begin{align}
{h_{k,n}} = \frac{{\left( {l + 1} \right){\theta _l}{\theta _c}{A_{\mathrm{p}}}}}{{2\pi d_{k,n}^2}}{\cos ^l} \left( {{\phi _n}} \right)\cos \left( {{\varphi _k}} \right)\mathbb{I}\left( {{\varphi _k}} \right), 
\end{align}
where $\theta _l$ and $\theta _c$ represent the conversion efficiency from light to current and the opposite, respectively; $l$ represents the Lambertian order; $\phi_n$ and ${{\varphi _k}}$ represent the radiance angle and incidence angle, respectively; ${d_{k,n}}$ represents the distance between user $k$ and the $n$th LED; $\mathbb{I}\left ( \varphi _k \right )$ represents an indicator function. 
Denote $\phi_{\mathrm{FOV}}$ as the field of view (FOV) of each solar panel. 
When ${{\varphi _k}}$ satisfies ${\left| {{\varphi _k}} \right| \le {\varphi _{{\mathrm{FOV}}}}}$, $\mathbb{I} \left( {{\varphi _k}} \right) = 1$, otherwise $\mathbb{I} \left( {{\varphi _k}} \right) = 0$; $A_{\mathrm{p}}$ represents the effective physical area of each solar panel, which is given by
\begin{align}
{A_{\mathrm{p}}} = \frac{{i_r^2}}{{{{\sin }^2}\left( {{\varphi _{{\mathrm{FOV}}}}} \right)}}{A_{{\mathrm{s}}}},
\end{align}
where $i_r$ denotes the refractive index and $A_{{\mathrm{s}}}$ denotes the solar panel's detector area.
\subsection{Transmitter Model}


The entire transmission process from the LEDs to all users is facilitated by 1-layer RSMA \cite{Mao_2022_rsma_survey}.
Specifically, at the transmitter, message $M_k$ for user $k$, $ \forall k \in {\mathcal{K}}$ is split into a common part $M_{k,c}$ and a private part $M_{k,p}$. 
The common parts of all $K$ users $\left\{ M_{k,c} \right\}_{k = 1}^K$ are combined into one common message ${M_c}$, i.e., $M_c = \left\{ M_{1,c}, \ldots ,M_{K,c} \right\}$. 
By employing a shared or private codebook, the common message $M_c$ or each private message $M_{k,p}$ can be encoded into a common stream $s_0$ or a private stream $s_k$, respectively.
 
Due to the LED characteristics \cite{S_Ma_2017}, these $K+1$ signals $\left\{ {{s_i}} \right\}_{i = 0}^K$ should satisfy  the peak amplitude requirement $\left| {{s_i}} \right| \le {A_i}$, with mean $\mathbb{E} \left \{ s_i \right \} =0 $ and variance $\mathbb{E} \left \{ s_i^2 \right \} =\varepsilon _i$, $\forall i \in {\mathcal{I}} = \left\{ {0,1, \ldots ,K} \right\}$.
Each stream $s_i$ is linearly precoded by the beamforming vector ${{\mathbf{p}}_i} = {\left[ {{p_{i,1}}, \ldots ,{p_{i,N}}} \right]^T} \in {{\mathbb{R}}^{N \times 1}}$, $  \forall i \in \mathcal{I}$.
The resulting transmit signal is 
\begin{align} \label{transmit_signal}
\mathbf{x} = \sum\limits_{i = 0}^K \mathbf{p}_i  s_i +  \mathbf{b},
\end{align}
where  ${\mathbf{b}} = {\left[ {b, \ldots ,b} \right]^T} \in {{\mathbb{R}}^{N \times 1}}$ represents the direct current (DC) bias vector which ensures the non-negativity of the transmit signal, i.e., $\mathbf{x}\geq\mathbf{0}$.
For each LED, its DC bias must meet the condition $b \ge {\mathrm{0}}$ \cite{Ma_2019_SLIPT_system}.
Given this non-negativity feature, the beamforming vectors $\{\mathbf{p}_i\}_{i=0}^K$ should satisfy  
\begin{align}
\sum\limits_{i = 0}^K {{A_i}\left| {{p_{i,n}}} \right|}  \le b, \forall n \in \mathcal{N}.
\end{align}

Furthermore, the transmit electrical power  $\{\mathbf{p}_i\}_{i=0}^K$ cannot exceed the transmit power budget $P_t$, i.e., $\mathbb{E} \{ \sum\nolimits_{i = 0}^K  \left \| \mathbf{p}_i  s_i  \right \|^2   \}  = \sum\nolimits_{i = 0}^K  \left \|  \mathbf{p}_i \right \|^2 \varepsilon _i \leq P_t$.

In addition to the aforementioned non-negative intensity constraint $\mathbf{x}\geq\mathbf{0}$, there are also illumination requirements within a VLC network. 
Specifically, let $I_{\mathrm{H}}$ and $I_{\mathrm{L}}$ denote the respective maximum and minimum permissible LED currents, the beamforming vectors $\{\mathbf{p}_i\}_{i=0}^K$ should also satisfy
 \begin{align} \label{dymatic_led_range}
\sum\limits_{i = 0}^K {A_i}{\mathbf{p}}_i^T{\mathbf{e}}_n + b \in \left [ I_{\mathrm{L}},I_{\mathrm{H}} \right ] ,\forall n \in \mathcal{N},
\end{align}
where ${\mathbf{e}}_n$ is an $N \times 1 $ unit vector with the $n$th element equal to 1 and all other elements equal to $0$.

Moreover, to meet the practical illumination requirements, dimming control is essential for regulating the average optical power \cite{Ma_2019_SLIPT_system}.
Here we denote $P_o^{\mathrm{ave}}$ and $\epsilon$ as the average optical power and dimming level, respectively. 
$P_o^{\mathrm{ave}}$ is directly determined by the DC bias and is given as $P_o^{\mathrm{ave}} = \mathbb{E}\left \{ \mathbf{x}  \right \}  =Nb$ \cite{Ma_2019_SLIPT_system}.
Then, the dimming level, defined as $\epsilon \in \left ( 0,1 \right ]$, can be expressed by
\begin{align} \label{dimming_control}
\epsilon  = \frac{P_o^{\mathrm{ave}}}{ P_o} =  \frac{Nb}{ P_o},
\end{align}
where $P_o$ denotes the maximum optical power.

By combining the dynamic current region \eqref{dymatic_led_range} and dimming control 
 \eqref{dimming_control}, we obtain an overall constraint for the optical power, which is given as
\begin{align} \label{optical_power_constriant}
 \sum_{i=0}^{K}A_i \mathbf{p}_i^T e_n \leq \min \left \{ \frac{\epsilon P_o}{N} -I_{\mathrm{L}},I_{\mathrm{H}}-\frac{\epsilon P_o}{N}  \right \} , \  \forall n \in \mathcal{N} .  
\end{align}


\subsection{Receiver Model}

As shown in Fig. \ref{system_model}, the receiver structure of each user is mainly based on the TS mode. 
Specifically, the TS mode enables SLIPT by dividing each time slot into two phases, one for  RSMA-aided ID module to receive information and the other  for the EH module to harvest energy. 
Let  $\theta \in \left ( 0,1 \right ] $ denote the fraction of time allocated to the ID phase. 
The remaining $1-\theta$ is allocated to the EH phase.

\textbf{Information Decoding Phase}: In  the RSMA-based ID phase, the signal received at user $k$ is given as
\begin{align}\label{receive_signal}
y_k = \mathbf{h}_{k}^T \mathbf{p}_0 s_0  + 
\sum_{i=1}^{K} \mathbf{h}_{k}^T\mathbf{p}_i s_i + \mathbf{h}_{k}^T \mathbf{b}  + n_k ,
\end{align}
where $n_k$ represents the Gaussian noise received at user $k$ with zero mean and variance $\sigma_k^2$. Based on \eqref{receive_signal}, we  then derive the lower bounds of the achievable rate for decoding the common stream $s_0$ and the private stream $s_k$ at user $k$, which is given as  \cite{S_Ma_2017}, 
\begin{subequations}
\begin{align}
& R_{k,\mathrm{c}}  = \frac{1}{2} \log_{2}{\frac{2\pi \sigma_k^2 + \sum\limits_{i=0}^{K}\left | \mathbf{h}_k^T\mathbf{p}_i  \right |^2 \tau_i }{2\pi \sigma_k^2 + 2\pi \sum\limits_{j=1}^{K}\left | \mathbf{h}_k^T\mathbf{p}_j  \right |^2 \varepsilon _j }}  ,\\
& R_{k,\mathrm{p}} = \frac{1}{2}{\log _2}\frac{{2\pi \sigma _k^2  +\sum\limits_{i = 1}^K {{{\left| {{\mathbf{h}}_k^T{{\mathbf{p}}_i}} \right|}^2}{\tau _i}} }}
{{2\pi  \sigma _k^2  + 2\pi \sum\limits_{j = 1,j \ne k}^K {{{\left| \mathbf{h}_k^T \mathbf{p}_j \right|}^2}{\varepsilon _j}}  }},
\end{align}
\end{subequations}
where ${\tau _i}   = {e^{1 + 2\left( {{\alpha _i} + {\gamma _i}{\varepsilon _i}} \right)}}$.
$\alpha _i$, $\gamma _i$, and $\varepsilon _i$ are the distribution parameter of signal $s_i$, as detailed in \cite{S_Ma_2017}.

Since only a portion $\theta$ of the time is allocated to the ID phase, the achievable rate lower bounds are scaled by $\theta$, given as
\begin{subequations}
\begin{align}
& R_{k,\mathrm{c}}^{\mathrm{TS}}  =  \theta R_{k,\mathrm{c}}, \\
& R_{k,\mathrm{p}}^{\mathrm{TS}}  =  \theta R_{k,\mathrm{p}}.
\end{align}    
\end{subequations}
To guarantee that all users can successfully decode $s_0$, the lower bound of the achievable common rate should satisfy $R_{\mathrm{c}} = \min_{k \in \mathcal{K} } \left \{ R_{k,\mathrm{c}}^{\mathrm{TS}} \right \} = \sum\nolimits_{k = 1}^K c_k $, where $c_k $ is the portion of $R_{\mathrm{c}} $ at user $k$.
The lower bound of the total achievable rate for TS mode at user $k$ is $ 
R_{k} =c_{k}  + R_{k,{\mathrm{p}}}^{\mathrm{TS}}, ~\forall k \in \mathcal{K}$.


\textbf{Energy Harvesting Phase}: Following \cite{Ma_2019_SLIPT_system}, the harvested energy is derived based on the nonlinear current-voltage feature of the solar panels, which is given as
\begin{align}
E_k(\mathbf{p}_i,\mathbf{b}) &=   \Pi\sum_{i=0}^{K}  \left | \mathbf{h}_k^T\mathbf{p}_i  \right |^2   \varepsilon_k \nonumber \\
& + \Gamma \left ( \frac{\kappa }{A_s} \mathbf{h}_k^T(a \mathbf{b} + z \mathbf{1}_N) + E_a\right ) ^2 \nonumber \\
& + \Gamma  \left(\ln \mu-1\right) \left (\frac{\kappa }{A_s} \mathbf{h}_k^T(a \mathbf{b} + z \mathbf{1}_N)+E_a \right ),
\end{align}
where $\mathbf{1}_N$ represents an $N \times 1$ vector where all elements are equal to $1$ and $\Pi$, $\Gamma$, $\mu$, $\kappa$, $a$, $z$, $E_a$ are all constants as detailed in \cite{Ma_2019_SLIPT_system}.

In the EH phase, we aim at maximizing the harvested energy and there is no information transmission, i.e., $\mathbf{p}_i = \mathbf{0}, \forall i \in \mathcal{I}$. Therefore, the DC bias is set to its maximum value, i.e., $b = I_{\mathrm{H}}$.
Thus, the harvested energy at user $k$ during phase 2 can be given by
\begin{align}
E^{\mathrm{TS}}_k   = (1-\theta) E_k (\mathbf{0},I_{\mathrm{H}} \mathbf{1}_N).
\end{align}  

\subsection{Problem Formulation}
In this work, we aim at maximizing the worst case rate among users for ID under the harvested energy constraint for each user.
The formulated optimization problem is given as
\begin{subequations}\label{problem_original_ts}
\begin{align}
\mathbf{\mathcal{P}_0 :\quad}  & \max_{\left \{ \mathbf{p}_i  \right \} _{i=0}^K,\left \{c_k \right \} _{k=1}^K,\theta}   \min_{k \in \mathcal{K} }\left \{ c_{k}  + R_{k,{\mathrm{p}}}^{\mathrm{TS}} \right \}  \label{ts_ini_obj}\\
\st \ \ \ &  \sum_{i=1}^{K}c_i \leq R_{k,c}^{\mathrm{TS}} ,\forall k \in \mathcal{K} ,  \label{ts_ori_c1}\\ 
&  c_k  \geq 0 , \forall k \in \mathcal{K}, \label{ts_ori_c11}\\ 
&  E_k^{\mathrm{TS}}\geq E_{th}  ,  \forall k \in \mathcal{K}, \label{ts_ori_c2} \\
&  0 < \theta \leq 1 \label{ts_ori_c3}, \\
&   \sum_{i=0}^{K}A_i \mathbf{p}_i^T e_n \leq \min\left \{ \frac{\epsilon P_o}{N} -I_{\mathrm{L}},I_{\mathrm{H}}-\frac{\epsilon P_o}{N}  \right \} , \  \forall n \in \mathcal{N} ,   \label{ts_ori_c4}\\ 
&   \sum_{i=0}^{K} \left \|  \mathbf{p}_i \right \|^2 \varepsilon _i \leq P_t , \label{ts_ori_c5}
\end{align}
\end{subequations}
where $E_{th}$ represents the harvested energy lower bound at each user.
 
\section{optimization framework}
Problem $\mathcal{P}_0$ is non-convex with coupled variables. 
In this section, we propose a constrained-concave-convex programming (CCCP)-based optimization algorithm to address this problem.

It is obvious that the TS parameter $\theta$ and the precoders $\left \{ \mathbf{p}_i  \right \} _{i=0}^K$ are coupled in the objective function \eqref{ts_ini_obj} and constraint \eqref{ts_ori_c1}.
To remove this coupling, we introduce additional slack variables $t$, $\{v_{k,\mathrm{c}}\}_{k=1}^K$, $\{v_{k,\mathrm{p}}\}_{k=1}^K$ and utilize an  bilinear function transformation \cite{mao_2020_mmf_coupling}, specifically $ ab = \frac{1}{4} \left (  a+b\right )^2 - \frac{1}{4} \left (  a-b\right )^2 $.
Problem $\mathbf{\mathcal{P}_0}$ is then equivalently transformed into the following problem
\begin{subequations}
\begin{align}
\mathbf{\mathcal{P}_1: \quad}  & \max_{\Omega}\ \ \    t \\
\st \ \ \ & R_{k,\mathrm{c}} \geq v_{k,\mathrm{c}},\ R_{k,\mathrm{p}} \geq v_{k,\mathrm{p}} ,\forall k \in \mathcal{K} ,\label{ts_p1_c1}\\
& \frac{1}{4}\left ( \theta  + v_{k,\mathrm{c}}  \right )^2 - \frac{1}{4}\left (  \theta  - v_{k,\mathrm{c}}   \right )^2   \geq \sum_{i=1}^{K} c_i , \forall k \in \mathcal{K},\label{ts_p1_c2}\\
& \frac{1}{4}\left ( \theta  + v_{k,\mathrm{p}}  \right )^2 - \frac{1}{4}\left (  \theta  - v_{k,\mathrm{p}}   \right )^2 \geq t - c_k , \forall k \in \mathcal{K} ,\label{ts_p1_c3}\\
& \eqref{ts_ori_c11}, \eqref{ts_ori_c2}, \eqref{ts_ori_c3}, \eqref{ts_ori_c4} ,\eqref{ts_ori_c5},\nonumber
\end{align}
\end{subequations}
where $ \Omega  = \left \{ \mathbf{p}_i,c_k ,v_{k,\mathrm{c}},v_{k,\mathrm{p}} ,\theta,t | \forall i \in \mathcal{I},\forall k \in \mathcal{K} \right \}$.
However, problem $\mathbf{\mathcal{P}_1}$ remains non-convex. To address this, we apply SDR and CCCP techniques to effectively solve problem $\mathbf{\mathcal{P}_1}$.

Specifically, we first employ the SDR technique and define $\mathbf{H}_k = \mathbf{h}_k \mathbf{h}_k^T$, $\forall k \in \mathcal{K}$, $\mathbf{P}_i = \mathbf{p}_i \mathbf{p}_i^T$, $\forall i \in \mathcal{I}$,   where $\mathrm{rank}(\mathbf{P}_i ) = 1$.
Constraint \eqref{ts_p1_c1} is then equivalently  transformed into
\begin{subequations}  
\begin{align}
& {\log _2}\left( 2\pi \sigma _k^2 +  \sum_{i=0}^{K}  \tau_i \mathrm{Tr}\left (  \mathbf{H}_k \mathbf{P}_i \right )    \right) \nonumber \\
& - {\log _2}\left( 2\pi \sigma _k^2 + 2\pi  \sum_{j=1}^{K} \varepsilon _j\mathrm{Tr}\left (  \mathbf{H}_k \mathbf{P}_j \right )  \right)\geq 2 v_{k,\mathrm{c}},\forall k \in {\mathcal{K}},\label{ts_p1_sdr1} \\
&  {\log _2}\left( 2\pi \sigma _k^2 +  \sum_{i=1}^{K}  \tau_i \mathrm{Tr}\left (  \mathbf{H}_k \mathbf{P}_i \right )    \right)\nonumber \\
& -{\log _2}\left( 2\pi \sigma _k^2 + 2\pi  \sum_{j=1,j \ne k}^{K} \varepsilon _j\mathrm{Tr}\left (  \mathbf{H}_k \mathbf{P}_j \right )  \right) \geq 2 v_{k,\mathrm{p}},\forall k \in {\mathcal{K}},\label{ts_p1_sdr2}\\
& \mathbf{P}_i \succeq \mathbf{0},\forall i \in {\mathcal{I}}, \label{ts_p1_sdr3}\\
& \mathrm{rank}\left ( \mathbf{P}_i  \right )   = 1, \ \forall  i \in \mathcal{I}. \label{ts_p1_sdr4}
\end{align}
\end{subequations}
Similarly, constraints \eqref{ts_ori_c4}, \eqref{ts_ori_c5} are transformed to:
\begin{subequations} \label{ts_power_constraint_sdr}
\begin{align}
&\sum\limits_{i = 0}^K {A_i^2{\mathrm{Tr}}\left( {{{\mathbf{P}}_i}{{\mathbf{e}}_n}{{\mathbf{e}}_n}^T} \right)}  \le \min\left\{ {{{\left( {\epsilon \frac{P_o}{N}  - {I_{\mathrm{L}}}} \right)}^2},} \right. \label{ts_p1_sdr5}\nonumber\\
& \quad \quad \quad \quad \quad  \quad \quad \quad \quad \quad \left. {{{\left( {{I_{\mathrm{H}}} -\epsilon \frac{P_o}{N}  } \right)}^2}} \right\},\forall n \in \mathcal{N},\\
&   \sum_{i=0}^{K} \mathrm{Tr}\left ( \mathbf{P}_i  \right )  \varepsilon _i \leq P_t . \label{ts_p1_sdr6}
\end{align}
\end{subequations}
\par
Furthermore, we observe that the left-hand sides of  constraints \eqref{ts_p1_sdr1}, \eqref{ts_p1_sdr2}, \eqref{ts_p1_c2}, and \eqref{ts_p1_c3} comprise both concave and convex terms, motivating us to apply CCCP.
Through this approach,  problem $\mathbf{\mathcal{P}_1}$ can be addressed iteratively until convergence is achieved, yielding high-quality sub-optimal solutions.
Specifically, at iteration $[m+1]$, the second terms in \eqref{ts_p1_sdr1} and \eqref{ts_p1_sdr2} are approximated by applying the first-order Taylor series expansion at a given point $\mathbf{P}_i^{\left [ m \right ] }$, $\forall i \in \mathcal{I}$ as
\begin{subequations} \label{cccp_rate1}
\begin{align}
& {\log _2}\left( 2\pi \sigma _k^2 +  \sum_{i=0}^{K}  \tau_i \mathrm{Tr}\left (  \mathbf{H}_k \mathbf{P}_i \right )    \right)   - F_{k,\mathrm{c} }^{\left [ m \right ] }\left ( \mathbf{P}_i   \right ) \geq 2 v_{k,\mathrm{c} },\label{cccp_1}\\
& {\log _2}\left( 2\pi \sigma _k^2 +  \sum_{i=1}^{K}  \tau_i \mathrm{Tr}\left (  \mathbf{H}_k \mathbf{P}_i \right )    \right)   - F_{k,\mathrm{p} }^{\left [ m \right ] }\left ( \mathbf{P}_i  \right ) \geq 2 v_{k,\mathrm{p} }, \label{cccp_2} 
 \end{align}
\end{subequations}
where $F_{k,\mathrm{p} }^{\left [ m \right ] }\left ( \mathbf{P}_i  \right ) $ and $F_{k,\mathrm{c} }^{\left [ m \right ] }\left ( \mathbf{P}_i  \right ) $ are  given by
\begin{subequations}
\begin{align}
& F_{k,\mathrm{p} }\left ( \mathbf{P}_i   \right ) = {\log _2}\left( 2\pi \sigma _k^2 + 2\pi  \sum\nolimits_{j=1,j \ne k}^{K} \varepsilon _j\mathrm{Tr}\left (  \mathbf{H}_k \mathbf{P}_j^{\left [ m \right ] } \right )  \right) \nonumber \\ 
&\quad  + \frac{  \sum _{j=1,j \ne k}^{K} 2\pi\varepsilon _j \mathrm{Tr}\left (  \mathbf{H}_k \left ( \mathbf{P}_j -\mathbf{P}_j^{\left [ m \right ] } \right )   \right )  }{\left ( 2\pi \sigma _k^2 + 2\pi  \sum_{j=1,j \ne k}^{K} \varepsilon _j\mathrm{Tr}\left (  \mathbf{H}_k \mathbf{P}_j^{\left [ m \right ] } \right )  \right ) \ln{2} } , \\
& F_{k,\mathrm{c} }\left ( \mathbf{P}_i   \right ) = {\log _2}\left( 2\pi \sigma _k^2 + 2\pi  \sum\nolimits_{j=1 }^{K} \varepsilon _j\mathrm{Tr}\left (  \mathbf{H}_k \mathbf{P}_j^{\left [ m \right ] } \right )  \right) \nonumber \\ 
& \quad \quad \quad+ \frac{  \sum _{j=1 }^{K} 2\pi\varepsilon _j \mathrm{Tr}\left (  \mathbf{H}_k \left ( \mathbf{P}_j -\mathbf{P}_j^{\left [ m \right ] } \right )   \right )  }{\left ( 2\pi \sigma _k^2 + 2\pi  \sum_{j=1}^{K} \varepsilon _j\mathrm{Tr}\left (  \mathbf{H}_k \mathbf{P}_j^{\left [ m \right ] } \right )  \right ) \ln{2} } .
\end{align}
\end{subequations}
Here $\mathbf{P}_i^{\left [ m \right ] }$, $\forall i \in \mathcal{I}$ is a feasible point attained from iteration $[m]$.

Similarly, the first terms of \eqref{ts_p1_c2} and \eqref{ts_p1_c3} are approximated as
\begin{subequations}\label{cccp_rate2}
\begin{align}
& G_{k,\mathrm{c}} ^{\left [ m \right ]}\left ( \theta,v_{k,\mathrm{c}} \right ) - \frac{1}{4}\left (  \theta  - v_{k,\mathrm{c}}   \right )^2   \geq \sum_{i=1}^{K} c_i \label{cccp_3} ,\\
& G_{k,\mathrm{p}} ^{\left [ m \right ]}\left ( \theta,v_{k,\mathrm{p}} \right ) - \frac{1}{4}\left (  \theta  - v_{k,\mathrm{p}}   \right )^2 \geq t - c_k ,\label{cccp_4} 
\end{align}
\end{subequations}
where $G_{k,\mathrm{c}}^{\left [ m \right ]}\left ( \theta,v_{k,\mathrm{c}} \right )$ and $G_{k,\mathrm{p}}^{\left [ m \right ]}\left ( \theta,v_{k,\mathrm{c}} \right )$ are
\begin{subequations}
\small
\label{eq:GkcGkp}
\begin{align}
 G_{k,\mathrm{c}} ^{\left [ m \right ]}\left ( \theta,v_{k,\mathrm{c}} \right ) = &\frac{1}{2} \left (\theta^{\left [ m \right ] } +  v_{k,\mathrm{c}}^{\left [ m \right ] }  \right )\left (\theta +  v_{k,\mathrm{c}} \right ) - \frac{1}{4} \left (\theta^{\left [ n \right ] } +  v_{k,\mathrm{c}}^{\left [ m \right ] }  \right ) ^2 ,  \\
G_{k,\mathrm{p}} ^{\left [ m \right ]}\left ( \theta,v_{k,\mathrm{p}} \right ) =  &\frac{1}{2} \left (\theta^{\left [ m \right ] } +  v_{k,\mathrm{p}}^{\left [ m \right ] }  \right )\left (\theta +  v_{k,\mathrm{p}} \right )- \frac{1}{4} \left (\theta^{\left [ n \right ] } +  v_{k,\mathrm{p}}^{\left [ m \right ] }  \right ) ^2.
\end{align}
\end{subequations}
$\theta^{\left [ m \right ] }$, $v_{k,\mathrm{c}}^{\left [ m \right ] }$ and $v_{k,\mathrm{p}}^{\left [ m \right ] }$, $\forall k \in \mathcal{K}$ in \eqref{eq:GkcGkp} represents a feasible point attained from iteration $[m]$.
\par 
So far, we have addressed all the non-convex constraints except for the rank-one constraint \eqref{ts_p1_sdr4} caused by SDR.
To handle it, we adopt a penalty method.
Specifically, we define $\mathbf{\mathbf{\xi}} _{i }^{\left [ m \right ] }\in {\mathbb{R}}^{N \times 1}$, $\forall i \in \mathcal{I}$ as the normalized eigenvectors associated with the principal eigenvalues of $\mathbf{P}_i^{\left [ m \right ] }$, $\forall i \in \mathcal{I}$.
Then we transform the rank-one constraint to $\sum_{i=0}^{K} \left ( \mathrm{Tr}\left ( \mathbf{P}_i  \right )- \left \| \mathbf{P}_i  \right \| _2   \right ) = 0$. 
We then incorporate the rank-one constraint into the objective function by adding the following penalty term:
\begin{align}
F_{\mathrm{penalty}} = \rho  \sum_{i=0}^{K} \left [ \mathrm{Tr}\left ( \mathbf{P}_i \right )  - \left ( \mathbf{\xi} _{i}^{\left [ m \right ] } \right )^T \mathbf{P}_i  \mathbf{\xi} _{i }^{\left [ m \right ] }\right ] ,
\end{align}
where $\rho$ is a negative penalty constant, which should be selected to guarantee that $F_{\mathrm{penalty}}$ approaches zero closely.

Therefore, we can solve problem $\mathbf{\mathcal{P}_0 }$ via a sequence of convex sub-problems.
Specifically, at iteration $[m+1]$,  based on the optimal solution $\mathbf{P}_i^{\left [ m \right ] },  \mathbf{\xi} _{i }^{\left [ m \right ] }$ obtained from the $[m]$th iteration, we solve the following sub-problem: 
\begin{subequations}
\begin{align}
\mathbf{\mathcal{P}_2 :\quad}  & \max_{\Omega} \ \ \   t +F_{\mathrm{penalty}} \\
\st \ \ \ &  \eqref{ts_ori_c11}  ,\eqref{ts_ori_c2} ,\eqref{ts_ori_c3} ,\eqref{ts_p1_sdr3},  \eqref{ts_power_constraint_sdr},\eqref{cccp_rate1},\eqref{cccp_rate2} . \nonumber
\end{align}  
\end{subequations}
The optimization procedure to solve problem $\mathbf{\mathcal{P}_0}$ is detailed in Algorithm 1.
\begin{algorithm}[htb]
	\caption{Proposed CCCP-based algorithm for solving $\mathbf{\mathcal{P}_0}$ }
	\begin{algorithmic}[1]
		\State \textbf{Input}: Initialize $m=0$, feasible $\mathbf{P}_i^{\left[m \right]}$, $\mathbf{\xi} _{i}^{\left [ m \right ] }$, $i \in \mathcal{I}$, define ${\ t ^{\left[m \right]}} \leftarrow 0$, the penalty factor $\rho$, convergence accuracy $\zeta$;
		\State \textbf{Repeat}
          \State \hspace*{0.5em} $m \leftarrow m + 1$;
          \State \hspace*{0.5em} Solve problem $\mathbf{\mathcal{P}_2 }$ with given $\mathbf{P}_i^{\left[m-1 \right]}$, $\mathbf{\xi} _{i }^{\left [ m-1 \right ] }$, $i \in \mathcal{I}$ 
             
           \Statex \hspace*{0.5em} and the corresponding solution is $\mathbf{P}_i^{*}$, $\mathbf{\xi} _{i }^{*}$, $i \in \mathcal{I}$. The 
          \Statex \hspace*{0.5em} optimal objective  value is obtained as $t^* + F_{\mathrm{penalty}}^*$ .
          \State \hspace*{0.5em} Update $ t ^{\left[ m \right]} \leftarrow {\ t ^*}$, $\mathbf{P}_i^{\left[ m \right]} \leftarrow {\mathbf{P}}_i^*$, $\mathbf{\xi} _{i }^{\left[ m \right]} \leftarrow \mathbf{\xi} _{i}^*$, $i \in \mathcal{I}$.    
           
		\State  \textbf{until} $ \left |  \left ( t^{\left [ m+1 \right ] } + F_{\mathrm{penalty}}^{\left [ m+1 \right ] }  \right ) - \left ( t^{\left [ m \right ] } + F_{\mathrm{penalty}}^{\left [ m \right ] } \right )  \right | < \zeta $;
        \State  \textbf{Output}: $\mathbf{p}_i^*= \mathrm{EVD} \left ( \mathbf{P}_i^* \right ) $, ${\ t ^*}$.
	\end{algorithmic} \label{algorithm_penalty}
\end{algorithm}
Specifically, in each iteration, we solve problem $\mathbf{\mathcal{P}_2}$ to update $\mathbf{P}_i^{ \left [ m \right ] }$, $\mathbf{\xi} _{i }^{\left [ m \right ] }$ and $t^{ \left [ m \right ] }$.
The iteration ends until the convergence of the objective function. $\zeta$ in Algorithm 1 denotes the error tolerance.
As $\mathbf{P}_i^*$ is guaranteed to be rank-one by the penalty term, we can then use EVD to recover the near-optimal beamforming vector $\mathbf{p}_i^*$ from $\mathbf{P}_i^*$.
As the interior-point method is used to address problem $\mathbf{\mathcal{P}_2}$, the worst-case computational complexity of Algorithm 1 as $\mathcal{O} \left (  \log{ \left ( \zeta^{-1} \right ) } \left [ KN^2  \right ]^{3.5} \right )$.

\section{numerical results}
In this section, we evaluate the performance of the proposed RSMA-aided multi-user MISO SLIPT.
The simulation setup considered in this section follows existing works \cite{S_Ma_2017,Ma_2019_SLIPT_system}.
Specifically, we consider a downlink VLC network deployed in a room with a size of $3$ m $\times$ $3$ m $\times$ $5$ m. 
The location in the room is modeled by a three-dimensional coordinate $(X, Y, Z)$, with one vertex of the room designated as the origin $(0, 0, 0)$.
There exists in total eight LEDs $(N=8)$ in the ceiling serving multiple users uniformly distributed in the receiving plane. 
The vertical height of all users is fixed as $1.7$ m, i.e., the location of user-$k$ is defined as ($x_k$, $y_k$, $1.7$).
Without loss of generality, the peak amplitude of signal, the variance of input signal, the average electrical noise power are equal for all users, i.e., $A = A_0 = A_1 = \cdots  = A_K,\ \varepsilon = \varepsilon_0   = \varepsilon_1 = \cdots  = \varepsilon_K$ and $\sigma^2 = \sigma^2_1 = \cdots  = \sigma^2_K$.
The locations of LEDs and users are specified in Table \ref{table_1}.
In Table \ref{table_2}, we further illustrate the detailed parameters of the considered setup.
In this work, we consider SDMA and NOMA-aided SLIPT \cite{2020_noma_slipt} as two benchmark schemes. 
It is worth noting that the proposed algorithm can be directly applied to address the MMF problem for both SDMA and NOMA-aided SLIPT since SDMA and NOMA are two special instances of RSMA \cite{Mao_2022_rsma_survey}.


\begin{table}[t!]
\centering
\caption{\normalsize LOCATIONS OF LEDS.}
\begin{tabular}{|c|c|c|c|}
\hline
       & Coordinate &        & Coordinate  \\
\hline
${\mathrm{LED }}\;1$ & (0.5,2.5,4.5)  &${\mathrm{LED }}\;2$ & (2.5,0.5,4.5)\\
  \hline
${\mathrm{LED }}\;3$ &(0.5,0.5,4.5)& ${\mathrm{LED }}\;4$ &(2.5,2.5,4.5)\\
  \hline
${\mathrm{LED }}\;5$ &  (0.5,1.5,4.5) & ${\mathrm{LED }}\;6$ &(2.5,1.5,4.5)\\
  \hline
${\mathrm{LED }}\;7$ &(1.5,0.5,4.5)&${\mathrm{LED }}\;8$  &  (1.5,2.5,4.5)\\
  \hline
\end{tabular}\label{table_1}
\end{table}
\begin{table}[t!]
	\caption{SUMMARY OF LED PARAMETERS AND THEIR VALUES.}
	\label{table_2}
	\centering
	\begin{tabular}{|m{3.5cm}|m{1cm}<{\centering}|m{2.2cm}<{\centering}|}
        \hline
		Peak amplitude of signals  & $A$ & $2\;{\mathrm{V}}$\\
		\hline
		Variance of signals & $\varepsilon$  & 1 \\
        \hline
		Receiver FOV & ${\varphi _{{\mathrm{FOV}}}}$ & ${60^ \circ }$ \\
		\hline
		LED emission semi-angle & ${\phi _{1/2}}$ & ${60^ \circ }$ \\
        \hline
		Detector area of solar panel & ${A_{{\mathrm{s}}}}$ & $10\;{\mathrm{c}}{{\mathrm{m}}^2}$ \\
		\hline
		Refractive index & $i_r$ & $1.5$ \\
		\hline
		Conversion efficiency  & ${\theta _l}$ & ${\mathrm{0}}{\mathrm{.54}}\left( {{\mathrm{A/W}}} \right)$\\
		\hline
		Average electrical noise power  & ${\sigma}^2$ & $-98.82\;\mathrm{dBm}$\\
		\hline
        dimming level & $\epsilon$ & $0.8$\\
		\hline
	\end{tabular}
\end{table}



\begin{figure*}
		\begin{center}
			\begin{minipage}{0.32\textwidth}
				\includegraphics[width=1\linewidth]{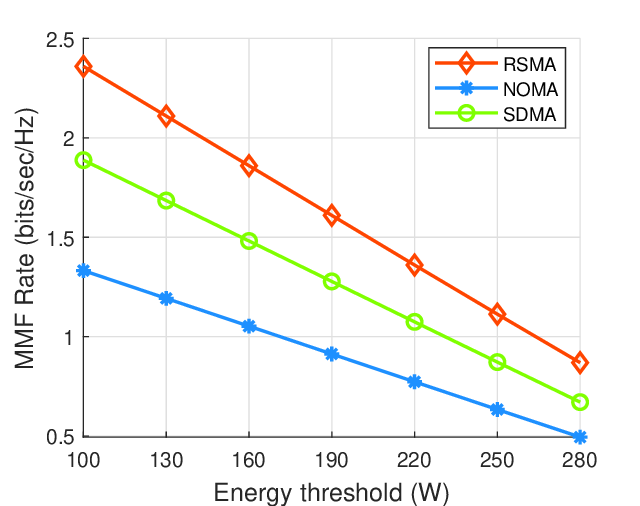}
		\caption{MMF rate (bits/sec/Hz) versus the energy threshold $E_{th}$ when $N=8$ LEDs serve $K=3$ users.}
		  \label{mmf_vs_Eth}
			\end{minipage}
   \hspace{1mm}
			\begin{minipage}{0.32\textwidth}
				\includegraphics[width=1\linewidth]{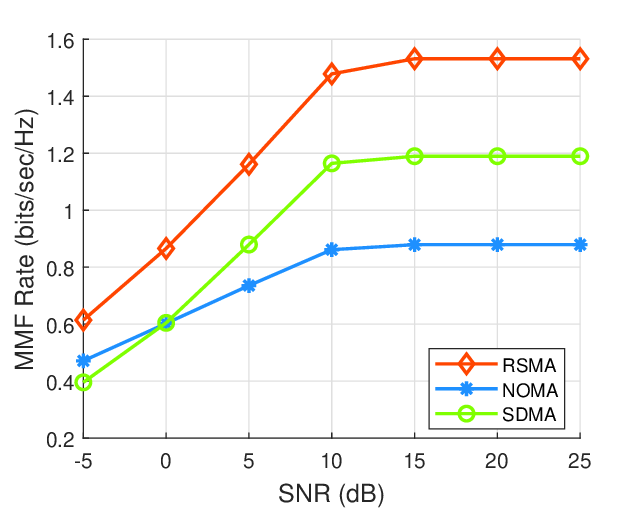}
		\caption{MMF rate (bits/sec/Hz) versus SNR when $N=8$ LEDs serve $K=3$ users.\\}
		 \label{mmf_vs_snr}
			\end{minipage}
   \hspace{1mm}
			\begin{minipage}{0.32\textwidth}
				\includegraphics[width=1\linewidth]{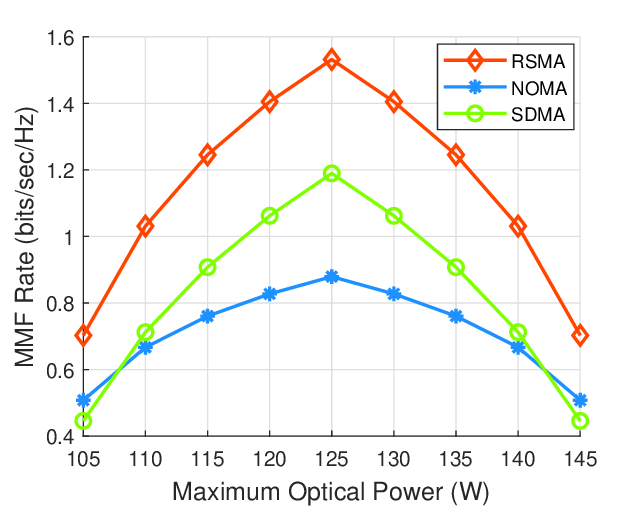}
		\caption{MMF rate (bits/sec/Hz) versus the maximum optical power $P_o$ when $N=8$ LEDs serve $K=3$ users.}
		 \label{mmf_vs_max_op_power}
			\end{minipage}
		\end{center}
  \vspace*{-0.5cm}
	\end{figure*}

We first  consider an underloaded scenario when $N=8$ LEDs serve $K=3$ users, SNR = 15 (dB), $\left [ I_{\mathrm{L}},I_{\mathrm{H}} \right ]  = \left [ 10,15 \right ] $  and $P_o = 125$ (W).
All results in this section consider the same $\left [ I_{\mathrm{L}},I_{\mathrm{H}} \right ] $ and $P_o$ unless otherwise noted.
Fig. \ref{mmf_vs_Eth} illustrates the MMF rate (bits/sec/Hz) versus the energy threshold $E_{th}$.
We observe that the MMF rates of all three MA schemes decrease as the energy threshold $E_{th}$ increases. RSMA outperforms other two schemes for all different energy thresholds.
Specifically, for low energy threshold, i.e., $E_{th} = 100$ (W), RSMA attains nearly $125 \%$ MMF rate gain over SDMA and $177 \%$ MMF rate gain over NOMA. 
For high energy threshold, i.e., $E_{th} = 280$ (W), RSMA attains nearly $130 \%$ MMF rate gain over SDMA and $176 \%$ MMF rate gain over NOMA. 

To investigate the influence of the transmit electrical power $P_t$ in \eqref{ts_ori_c5} to the performance, we illustrate the MMF rate versus the transmit SNR (dB) when $E_{th} = 200$ (W) in Fig. \ref{mmf_vs_snr}.
It is easy to observe that the MMF rates of all three MA schemes increase with SNR, while RSMA outperforms other two schemes for all different SNR.
Specifically, for low SNR, i.e., SNR = $-5$ dB, RSMA attains nearly $ 155 \%$ MMF rate gain over SDMA and $ 130 \%$ MMF rate gain over NOMA. 
For high energy threshold, i.e., SNR = $25$ dB, RSMA attains nearly $ 129\%$ MMF rate gain over SDMA and $174 \%$ MMF rate gain over NOMA. 
Meanwhile, the MMF rate performance for all three MA schemes is limited in the high SNR regime, i.e., SNR $\in \left [ 15,25 \right ] $ (dB). This is primary due to the limitations imposed by the optical power constraint \eqref{ts_ori_c4}.


To further investigate the effect of optical power constraint \eqref{ts_ori_c4}, we illustrate the MMF rate versus the maximum optical power $P_o$ when $E_{th} = 200$ (W) in Fig. \ref{mmf_vs_max_op_power}.
From \eqref{dimming_control}, we can observe that the maximum optical power directly affects the DC bias $b$ and constraint \eqref{ts_ori_c4}.
Specifically, the optimal MMF rate performance is reached when the maximum optical power is 
\begin{equation}
    P_o^{\star} =  \frac{N\left ( I_{\mathrm{L}}+I_{\mathrm{H}} \right ) }{2 \epsilon}.
\end{equation}
Furthermore, it is also shown that RSMA is superior to other two schemes across all different maximum optical power.

\section{conclusion}
In this paper, we introduce RSMA into a multi-user VLC SLIPT system where all users require to decode the intended information and harvest energy based on the time-splitting mode. 
We also propose a beamforming optimization algorithm to address the formulated MMF rate maximization problem subject to the energy harvesting constraint at each user. 
Specifically, we propose a CCCP-based algorithm, which exploits a bilinear function transformation, SDR, CCCP, and a penalty-based method to handle the non-convex constraints and objective functions. 
Numerical results show that the proposed RSMA-aided SLIPT shows significant performance gain over the conventional SDMA and NOMA. We therefore draw the conclusion that RSMA is a promising and powerful MA scheme for SLIPT systems.

\bibliographystyle{IEEE-unsorted}
\bibliography{VLC+SLIPT+RSMA_reference}

\begin{thebibliography}{10}

\bibitem{Dia_2018_SLIPT}
P.~D. Diamantoulakis, et~al.,
\newblock ``Simultaneous lightwave information and power transfer ({SLIPT}),''
\newblock {\em IEEE Trans.Green Commun. Netw.}, vol.~2, no.~3, pp.~764--773, 2018.

\bibitem{Mao_2022_rsma_survey}
Y.~Mao, et~al.,
\newblock ``Rate-splitting multiple access: Fundamentals, survey, and future research trends,''
\newblock {\em IEEE Commun. Surv. Tutorials}, vol.~24, no.~4, pp.~2073--2126, 2022.

\bibitem{Mao_EURASIP_2018}
Y.~Mao, et~al.,
\newblock ``Rate-splitting multiple access for downlink communication systems: {B}ridging, generalizing, and outperforming {SDMA} and {NOMA},''
\newblock {\em EURASIP J. Wireless Commun. Netw.}, pp. 1--54, May 2018.

\bibitem{Naser_2022}
S.~Naser, et~al.,
\newblock ``Coordinated beamforming design for multi-user multi-cell {MIMO} {VLC} networks,''
\newblock {\em IEEE Photonics J.}, vol.~14, no.~3, pp.~1--10, 2022.

\bibitem{Tao_ICC_2020}
S.~{Tao}, et~al.,
\newblock ``One-layer rate-splitting multiple access with benefits over power-domain {NOMA} in indoor multi-cell visible light communication networks,''
\newblock in {\em Proc. IEEE Int. Conf. Commun. China}, pp. 1--7, Jun. 2020.

\bibitem{2024_ris_vlc_rsma}
O.~Maraqa, et~al.,
\newblock ``Optical {STAR-RIS}-aided {VLC} systems: {RSMA} versus {NOMA},''
\newblock {\em IEEE Open J. Commun. Soc.}, vol.~5, pp.~430--441, 2024.

\bibitem{2024_Girdher_rsma_slipt}
A.~Girdher, et~al.,
\newblock ``Fairness-aware energy harvesting in {RSMA}-aided {MU-MISO} {VLC} network,''
\newblock {\em IEEE Commun. Lett.}, vol.~28, no.~5, pp.~1062--1066, 2024.

\bibitem{2024_guo_slipt_rsma}
Y.~Guo, et~al.,
\newblock ``Max-min fairness in rate-splitting multiple access-based {VLC} networks with {SLIPT},''
\newblock {\em IEEE Internet Things J.}, pp. 1--1, 2024.

\bibitem{2019_mao_rsma_swipt}
Y.~Mao, et~al.,
\newblock ``Rate-splitting for multi-user multi-antenna wireless information and power transfer,''
\newblock in {\em Proc. IEEE Int. Workshop Signal Process. Adv. Wireless Commun. (SPAWC)}, pp. 1--5, 2019.

\bibitem{S_Ma_2017}
S.~Ma, et~al.,
\newblock ``Achievable rate with closed-form for {SISO} channel and broadcast channel in visible light communication networks,''
\newblock {\em J. Lightw. Technol.}, vol.~35, no.~14, pp.~2778--2787, Jul. 2017.

\bibitem{wang_2015_solar_panel_design}
Z.~Wang, et~al.,
\newblock ``On the design of a solar-panel receiver for optical wireless communications with simultaneous energy harvesting,''
\newblock {\em IEEE J. Sel. Areas Commun.}, vol.~33, no.~8, pp.~1612--1623, 2015.

\bibitem{Ma_2019_SLIPT_system}
S.~Ma, et~al.,
\newblock ``Simultaneous lightwave information and power transfer in visible light communication systems,''
\newblock {\em IEEE Trans. Wireless Commun.}, vol.~18, no.~12, pp.~5818--5830, 2019.

\bibitem{mao_2020_mmf_coupling}
Y.~Mao, et~al.,
\newblock ``Max-min fairness of {K}-user cooperative rate-splitting in {MISO} broadcast channel with user relaying,''
\newblock {\em IEEE Trans. Wireless Commun.}, vol.~19, no.~10, pp.~6362--6376, 2020.

\bibitem{2020_noma_slipt}
X.~Liu, et~al.,
\newblock ``Beamforming design for secure {MISO} visible light communication networks with {SLIPT},''
\newblock {\em IEEE Trans. Commun.}, vol.~68, no.~12, pp.~7795--7809, 2020.

\end{thebibliography}

\end{document}